\documentclass[aps,prd,twocolumn,showpacs,floatfix,preprintnumbers,amsmath,amssymb,nofootinbib]{revtex4-1}
\input epsf
\usepackage{graphicx}

\newcommand{\beq}{\begin{equation}}
\newcommand{\eeq}{\end{equation}}
\newcommand{\barr}{\begin{eqnarray}}
\newcommand{\earr}{\end{eqnarray}}

\newcommand{\rme}{\textrm{e}}

\newcommand{\rmd}{\textrm{d}}

\newcommand{\A}{\mathcal{A}}
\newcommand{\B}{\mathcal{B}}
\newcommand{\R}{\mathcal{R}}

\newcommand{\Tm}{T_{\rm m}}
\newcommand{\Tr}{T_{\rm r}}

\newcommand{\apjl}{Astrophys. J. Lett.}
\newcommand{\aap}{Astron. Astrophys.}
\newcommand{\apjs}{Astrophys. J. Suppl. Ser.}

\newcommand{\mnras}{Mon. Not. R. Astron. Soc.}

\newcommand{\memras}{Mem. R. Astr. Soc. }

\newcommand{\lsim}{\mathrel{\hbox{\rlap{\lower.55ex\hbox{$\sim$}} \kern-.3em \raise.4ex \hbox{$<$}}}}
\newcommand{\gsim}{\mathrel{\hbox{\rlap{\lower.55ex\hbox{$\sim$}} \kern-.3em \raise.4ex \hbox{$>$}}}}

\begin{document}

\title{Ultrafast effective multi-level atom method for primordial hydrogen recombination}
\author{Yacine Ali-Ha\"imoud and Christopher M. Hirata} 
\affiliation{California Institute of Technology, Mail Code 350-17, Pasadena, CA 91125}
\date{\today}

\begin{abstract}
Cosmological hydrogen recombination has recently been the subject of renewed attention because of its importance for predicting the power spectrum of cosmic microwave background 
anisotropies. It has become clear that it is necessary to account for a large number $n \gtrsim 100$ of energy shells of the hydrogen atom, separately following the angular momentum substates 
in order to obtain sufficiently accurate recombination histories.  However, the multi-level atom codes that follow the 
populations of all these levels are 
computationally expensive, limiting recent analyses to only a few points in parameter space. In this 
paper, we present a new method for solving the multi-level atom recombination problem, which splits the problem into a computationally expensive atomic physics component that is independent of the 
cosmology, and an ultrafast cosmological evolution component. The atomic physics component follows the network of bound-bound and bound-free transitions among excited states and computes the resulting effective transition rates for the small set of ``interface'' states radiatively connected to the ground state. The cosmological evolution component only follows the populations of the interface states. By pre-tabulating the effective rates, we can reduce the recurring cost of multi-level atom calculations by more than 5 orders 
of magnitude.  The resulting code is fast enough for inclusion in Markov Chain Monte Carlo parameter estimation algorithms. It does not yet include the radiative transfer or high-$n$ two-photon processes 
considered in some recent papers.  Further work on analytic treatments for these effects will be required in order to produce a recombination code usable for {\slshape Planck} data analysis.
\end{abstract}


\maketitle

\section{Introduction}

The advent of high precision cosmic microwave background (CMB) experiments, such as \emph{Planck} \cite{Planck}, has recently motivated several authors to revisit the theory of cosmological 
recombination pioneered by Peebles \cite{Peebles} and Zeldovich et al. \cite{Zeldovich_et_al} in the 1960s. The free electron fraction as a function of redshift $x_e(z)$ is one of the major 
theoretical uncertainties in the prediction of the CMB temperature and polarization anisotropy power spectra \cite{Hu1995, Lewis06, Chluba10_uncertainties}. To obtain a recombination history accurate to 
the percent level, it is necessary to account for a high number of excited states of hydrogen, up to a principal quantum number $n_{\max} = \mathcal{O}(100)$ \cite{Recfast_long, 
Recfast_short}. The desired sub-percent accuracy can only be reached when explicitly resolving the out-of-equilibrium angular momentum substates, which requires 
the multi-level atom (MLA) codes to follow
$N_{\rm level} = n_{\max}(n_{\max}+1)/2$ individual states.  Moreover, the ordinary differential equations (ODEs) describing the level populations are stiff, requiring the 
solution of large $N_{\rm level}\times N_{\rm level}$ systems of equations at each integration time step.  This problem has been solved by several authors \cite{CRMS07, Grin_Hirata, Chluba_Vasil}, but each of these codes takes hours to days to run.

Eventually, it is necessary to be 
able to produce not only accurate but also {\em fast} recombination histories, to be included in Markov Chain Monte Carlo (MCMC) codes for cosmological parameter estimation.  The MCMC requires CMB power 
spectra (and hence recombination histories) to be generated at each proposed point in cosmological parameter space, with a typical chain sampling ${\cal O}(10^5)$ points \cite{WMAP7p}.  Furthermore, 
dozens of MCMCs are often run with different combinations of observational constraints and different parameter spaces.
This makes it impractical to 
include recombination codes that run for more than a few seconds in the MCMC.
One solution is to precompute 
recombination histories $x_e(z| H_0, T_{\rm CMB}, \Omega_m h^2, \Omega_b h^2, Y_{\rm He}, N_{\nu})$ on a grid of cosmological parameters, and then use elaborate interpolation algorithms to evaluate the 
recombination history for any cosmology \cite{RICO}, or to construct fitting functions \cite{Recfast_short, recfastimproved}.  However, such procedures need to be re-trained every time additional 
parameters are added, and are rather unsatisfying regarding their physical significance. 

In this work we present a new method of solution for the recombination problem, perfectly equivalent to the standard MLA method, but much more efficient computationally.  The basic idea is 
that the vast majority of the excited hydrogen levels are populated and depopulated only by \emph{optically thin} radiative transitions (bound-bound and bound-free) in a bath of thermal photons; we show that their effect can be ``integrated out'' leaving 
only a few functions of the matter and radiation temperatures $\Tm$ and $\Tr$ (this list would include the free electron density $n_e$ if 
we incorporated collisions), which can be pretabulated.  In an 
actual call to the recombination code from an MCMC, it is then only necessary to solve an \emph{effective} MLA (hereafter, EMLA) with a smaller number of levels (perhaps only 3: $2s$, $2p$ and $3p$), which eliminates the 
computationally difficult $N_{\rm level}\times N_{\rm level}$ 
system solution in the traditional MLA.  [The idea is similar in spirit to the line-of-sight integral method for the 
computation of the CMB power spectrum \cite{SZ.LOSI}, which eliminated a large number of independent variables from the cosmological perturbation theory system of ODEs (the high-order moments of the 
radiation field, $\Theta_\ell$ for $\ell\gg1$) in favor of 
pretabulated spherical Bessel functions.]  Our method achieves a speed-up of the recombination calculation by 5 to 6 orders of magnitude.

We note that our method only eliminates the computational complexity associated with the high-$n$ excited states and does not include continuum processes and radiative transfer effects 
that have been studied by previous authors \cite{DG05, CS06, Kholupenko06, WS07, CS08, Hirata_2photon, GD08, Hirata_Forbes, CS09a, CS09b, CS09c}.  However, we note that there has been much progress in 
analytic 
treatments of 
these effects \cite{Hirata_2photon, GD08, Hirata_Forbes}; ultimately, we expect to improve these analytic treatments and graft them (and an analytic treatment of helium recombination 
\cite{He1,He2,He3,HeKIV,HeRCS}) on to 
the 
ultrafast EMLA code described herein to yield a recombination code 
that is accurate enough for {\slshape Planck} data analysis.

This paper is organized as follows. In Section \ref{section:rates} we review the general picture of hydrogen recombination, and the bound-bound and bound-free transition rates involved in the 
calculation. In Section \ref{section:std MLA} we describe the standard MLA method. We present our new EMLA method in Section \ref{section:new method} and demonstrate its equivalence with the standard MLA formulation. We describe our numerical implementation and results in Section \ref{section:results}, and conclude in Section \ref{section:conclusion}. Appendix \ref{app:invertibility} is dedicated to demonstrating the invertibility of the system defining the EMLA equations. Appendix \ref{app:complementarity} proves a complementarity relation between effective transition probabilities. We prove detailed balance relations between effective transition rates in Appendix \ref{app:db}. Appendix \ref{appendix:post-saha} exposes the post-saha approximation we use at early times when computing recombination histories.

\section{Bound-bound and bound-free transition rates} \label{section:rates}


The evolution of the free electron fraction is governed by the network of transitions between bound states of hydrogen as well as recombination and photoionization rates. Before giving detailed 
expressions for these rates, let us first outline the general picture of the process of recombination.

It has long been known that direct recombinations to the ground state are ineffective for recombination \cite{Peebles, Zeldovich_et_al}, since the resulting emitted photons are immediately reabsorbed by 
hydrogen in 
the ground state, as soon as the neutral fraction is higher than $\sim 10^{-9}$. Electrons and protons can efficiently combine only to form hydrogen in excited states. The minute amount of excited 
hydrogen at all relevant times during cosmological recombination is not sufficient to distort the blackbody radiation field near the ionization thresholds of the excited states. Recombination to the 
excited states is therefore a thermal process: it depend on the matter temperature $\Tm$ which characterizes the free electrons and protons velocity distribution, and also on the radiation temperature 
$\Tr$, since the abundant low-energy thermal photons can cause stimulated recombinations. Photoionization rates from excited states only depend on the radiation temperature since they do not involve free 
electrons in the initial state.

Transitions between bound excited states may be radiative or collisional. Radiative transition rates are well known and depend only on the radiation temperature characterizing the blackbody radiation 
field, undistorted in the vicinity of the optically thin lines from the Balmer series and beyond. Collisional transition rates are much less precisely known, but only depend on the matter temperature and 
the abundance of charged particles causing the transitions (i.e. free electrons and free protons, which, once helium has recombined, have the same abundance $n_e = n_p$ due to charge neutrality).

Finally, some of the excited states can radiatively decay to the ground state. The most obvious route to the ground state is through Lyman transitions from the $p$ states. However, due to the very high 
optical depth of these transitions, emitted Lyman photons are immediately reabsorbed by hydrogen atoms in their ground state. This ``bottleneck'' can only be bypassed by the systematic redshifting of 
photons, which can escape re-absorption once their frequency is far enough below the resonant frequency of the line. The relevant transition rate in this case is the \emph{net} decay rate to the ground 
state, which is a statistical average accounting for the very small escape probability of Lyman photons. Two-photon transitions are usually much slower than single-photon transitions. However, the rate of two-photon decays from the metastable $2s$ state is comparable to the net decay rate in the highly self-absorbed Lyman transitions, and this process should therefore be included in a recombination calculation \cite{Peebles, 
Zeldovich_et_al}.

We now give explicit expressions for the bound-bound and bound-free rates dicussed above. Subscripts $nl$ refer to the bound state of principal quantum number $n$ and angular momentum quantum number $l$. 
We denote $\alpha_{\rm fs}$ the fine structure constant, $\mu_e \equiv m_em_p/(m_e + m_p)$ the reduced mass of the electron-proton system, $E_I$ the ionization energy of hydrogen, and $E_n \equiv 
-E_In^{-2}$
the energy of the $n^{\rm th}$ shell. Finally, we denote by $f_{\rm BB}(E, \Tr) \equiv ( \rme^{E/\Tr} - 1 )^{-1}$ the photon occupation number at 
energy $E$ in the blackbody radiation field at temperature $\Tr$.

\subsection{Recombination to and photoionization from the excited states}

The recombination coefficient to the excited state $nl$, including stimulated recombinations, is denoted $\alpha_{nl}(\Tm, \Tr)$ (it has units of cm$^{3}$ s$^{-1}$). The photoionization rate per atom in 
the state $nl$ is denoted $\beta_{nl}(\Tr)$. Both can be expressed in terms of the bound-free radial matrix elements $g(n,l,\kappa,l')$ \cite{Burgess}. Defining
\barr
\gamma_{nl}(\kappa) &\equiv& \frac2{3n^2} \alpha_{\rm fs}^3 \frac{E_I}{h} (1 + n^2 \kappa^2)^3 \nonumber \\
&&\times \sum_{l' = l\pm1} \max(l, l') g(n,l,\kappa,l')^2,
\earr
where $\kappa$ denotes the momentum of the outgoing electron in units of $\hbar/a_0$ (where $a_0$ is the reduced-mass Bohr radius),
the recombination coefficient is given by \cite{Burgess}:
\barr
\alpha_{nl}(\Tm, \Tr) &=&\frac{ h^3}{(2 \pi \mu_e T_m)^{3/2}}
\nonumber \\ &&
\times \int_0^{+\infty}\rme^{-E_I \kappa^2 / \Tm}\gamma_{nl}(\kappa)
\nonumber \\ && \times
 \left[1 +  f_{\rm BB}\left(E_{\kappa n},\Tr\right)\right]\rmd (\kappa^2),
\label{eq:alpha}
\earr
where $E_{\kappa n} \equiv E_I(\kappa^2 + n^{-2})$. The photoionization rate only depends on the radiation temperature and can be obtained by detailed balance considerations from the recombination 
coefficient: 
\beq
\beta_{nl}(\Tr) = \frac{(2 \pi \mu_e T_r)^{3/2}}{(2 l +1)h^3} \rme^{E_n /\Tr } \alpha_{nl}(\Tm = \Tr, \Tr).
\label{eq:beta-db}
\eeq

\subsection{Transitions between excited states}

We denote $R_{nl\rightarrow n'l'}$ the transition rate from the excited state $nl$ to the excited state $n'l'$. It has units of sec$^{-1}$ per atom in the initial state. Transitions among excited states 
can be either radiative or collisional: 
\beq
R_{nl\rightarrow n'l'} = R_{nl\rightarrow n'l'}^{\rm rad}(\Tr) + R_{nl\rightarrow n'l'}^{\rm coll}(\Tm, n_e),
\eeq 
where $n_e = n_p$ is the abundance of free electrons or free protons.  In this paper, we follow exclusively the radiative rates.  These are given by
\barr
R_{nl\rightarrow n'l'}^{\rm rad} = \Bigg{\{} \begin{array} {ccc} A_{nl,n'l'}\left[1 + f_{\rm BB}(E_{nn'}, \Tr)\right]& & E_n > E_{n'}\\[10pt]
 \frac{g_{l'}}{g_l} \rme^{-E_{n'n}/\Tr} R_{n' l'\rightarrow nl}^{\rm rad}& & E_n < E_{n'},
\end{array}\label{eq:Rab rad}
\earr
where $E_{nn'} \equiv E_n - E_{n'}$ is the energy difference between the excited levels, $g_l \equiv 2 l +1$ is the degeneracy of the state $nl$, and $A_{nl,n'l'}$ is the Einstein $A$-coefficient for the 
$nl \rightarrow n'l'$ transition, which may be obtained from the radial matrix element $R_{n'l'}^{nl}$\cite{Bethe}:
\beq
A_{nl,n'l'} = \frac{2\pi}3 \alpha_{\rm fs}^3 \frac{E_I}{h}\left(\frac1{n'^2}-\frac1{n^2}\right)^3 \frac{\max(l, l')}{2l +1}|R_{n'l'}^{nl}|^2.
\eeq


\subsection{Transitions to the ground state}

Finally, the ground state population $x_{1s} \approx 1 - x_e$ evolves due to transitions from and into the $np$ and $2s$ states (two-photon transitions from higher energy states are dominated by ``1 + 
1''  photon decays, already accounted for). Photons emitted in the Lyman lines are very likely to be immediately reabsorbed, and the only meaningful quantity for these transitions is the \emph{net} decay 
rate in the line, which is a statistical average over a large number of atoms, and accounts for the very low escape probability of a photon emitted in the line. In the Sobolev approximation 
\cite{Sobolev} with optical depth $\tau_{np,1s}\gg1$, the net 
decay rate in the $np\rightarrow 1s$ transition is:
\barr
\dot{x}_{1s}\big{|}_{np} = - \dot{x}_{np}\big{|}_{1s} &=& \frac{A_{np,1s}}{\tau_{np,1s}}\left(x_{np} - 3 x_{1s} f_{np}^+\right) \nonumber \\
&=& \frac{8 \pi H}{3 \lambda_n^3 n_{\rm H} x_{1s}}\left(x_{np} - 3 x_{1s} f_{np}^+\right),
\label{eq:dot.x_np}
\earr
where $\lambda_n \equiv h c /E_{n1}$ is the transition wavelength, and $f_{np}^+$ is the photon occupation 
number at 
the blue side of the corresponding Ly-$n$ line.  In this paper, we will take $f_{np}^+ = f_{\rm BB}(E_{n1}, \Tr)$, i.e. assume the incoming radiation on the blue side of the line has a blackbody spectrum.  This 
assumption is actually violated due to feedback from higher-frequency Lyman lines (e.g. radiation escaping from Ly$\beta$ can redshift into Ly$\alpha$) \cite{CS07, He1, Kholupenko_Deuterium}; while 
our formalism is general 
enough to incorporate different $f_{np}^+$, we have not yet implemented this in our code.

The $2s$ state cannot decay to the ground state through a radiatively allowed transition. This decay is however possible with a two-photon emission, which, although slow, is comparable in efficiency to 
the highly self-absorbed Lyman transitions. The simplest expression for the net $2s\rightarrow 1s$ two-photon decay rate is:
\barr
\dot{x}_{1s}\big{|}_{2s} = - \dot{x}_{2s}\big{|}_{1s} = \Lambda_{2s1s}\left(x_{2s} - x_{1s} \rme^{-E_2/\Tr}\right),
\label{eq:dot.x_2s}
\earr
where $\Lambda_{2s1s} \approx 8.22$ s$^{-1}$ is the total $2s\rightarrow 1s$ two-photon decay rate \cite{2photonrate}.

In each case, we denote the {\em net} downward rate in the $i\rightarrow 1s$ transition, where $i \in \{2s, 2p, 3p, ...\}$:
\beq
\dot{x}_{1s}\big{|}_{i} = - \dot{x}_i\big{|}_{1s} = x_i \tilde{R}_{i\rightarrow 1s} - x_{1s}\tilde{R}_{1s\rightarrow i},
\label{eq:xddown}
\eeq
where the rates $\tilde R$ depend on atomic physics, $\Tr$, and the optical depths in the Lyman lines.

Both the Sobolev approximation for the $np \rightarrow 1s$ transitions Eq.~(\ref{eq:dot.x_np}), and the simple expression Eq.~(\ref{eq:dot.x_2s}) for 
the net $2s \rightarrow 1s$ two-photon decay do not account for subtle yet important radiative transfer effects. An accurate recombination calculation 
should account for time-dependent effects in Ly$\alpha$ \cite{Hirata_2photon, CS09a}, a suite of two-photon continuum processes
\cite{CS06, Kholupenko06, Hirata_2photon, CS09b}, and resonant scattering in Ly$\alpha$ \cite{Hirata_Forbes, CS09c}.  These are not included in the present code and we plan to add them in the 
future using analytic treatments.

\section{The standard MLA method} \label{section:std MLA}

Although the standard MLA formulation does not make this distinction, we cast the excited states of hydrogen into two categories. On the one hand, most excited states are not directly radiatively 
connected to the ground state. We call these states ``interior'' states and denote $X_K$ the fractional abundance of hydrogen in the interior state $K \in \{3s, 3d, 4s, 4d, 4f, 5s, ...\}$. On the other hand, the $2s$ and $np$ states ($n\geq 2$) 
are directly radiatively connected with the ground state. We call these states ``interface'' states and denote $x_i$ the fractional abundance of hydrogen in the interface state $i \in \{2s, 2p, 3p ,...\}$. 

In the standard MLA formulation, the free electron fraction $x_e(z)$ is evolved by solving the hierarchy of coupled differential equations: for the 
interior states,
\barr
\dot{X}_K &=& x_e^2 n_{\rm H} \alpha_K + \sum_{L} X_L R_{L \rightarrow K} + \sum_j x_j R_{j \rightarrow K}  \nonumber\\
 &-& X_K \Bigl(\beta_K + \sum_L R_{K \rightarrow L} + \sum_j R_{K \rightarrow j} \Bigr);
\label{eq:dot.X_K}
\earr
for the interface states,
\barr
\dot{x}_i &=&  x_e^2 n_{\rm H} \alpha_i + \sum_{L} X_L R_{L \rightarrow i} +  \sum_j x_j R_{j \rightarrow i} + x_{1s} \tilde{R}_{1s\rightarrow i}\nonumber\\
 &&- x_i \Bigl( \beta_i + \sum_L R_{i \rightarrow L} + \sum_j R_{i \rightarrow j} + \tilde{R}_{i \rightarrow 1s}\Bigr); 
\label{eq:dot.x_i}
\earr
and for the free electrons and ground state,
\beq
\dot{x}_e = -\dot{x}_{1s} = x_{1s}\sum_i \tilde{R}_{1s\rightarrow i} - \sum_i x_i \tilde{R}_{i \rightarrow 1s}.\label{eq:dot.x_e}
\eeq
The radiative rates between excited states are many orders of magnitude larger than the rate at which recombination proceeds, which is of the order of 
the Hubble rate. Even the relatively small net rates out of the interface states ($\Lambda_{2s,1s}$ and $A_{2p,1s}/\tau_{2p,1s}$) are still more than 12 orders of magnitude larger than the Hubble rate. The populations of the excited states can therefore be obtained to high accuracy in the steady-state approximation 
(this approximation is ubiquitous in many problems and has long been used in the context of cosmological recombination \cite{Peebles, Hirata_2photon, Grin_Hirata}, where its accuracy has been tested explicitly \cite{Chluba_Vasil}).  
Setting $\dot{X}_K$ and $\dot{x}_i$ to zero in Eqs.~(\ref{eq:dot.X_K}) and (\ref{eq:dot.x_i}), we see that the problem amounts to first solve a system of 
linear algebraic equations for the $X_K, x_i$, with an inhomogeneous term depending on $x_e$, and then use the populations $x_i$ in 
Eq.~(\ref{eq:dot.x_e}) to evolve the free electron fraction. The 
solution of the system of equations (\ref{eq:dot.X_K}), (\ref{eq:dot.x_i}) needs to 
be done at \emph{every time step}, since the inhomogeneous term of the equation depends on the ionization history, which explicitly depends on time as well as on the cosmological parameters. Recent work 
\cite{Grin_Hirata, Chluba_Vasil} has shown that to compute sufficiently accurate recombination histories, one needs to account for excited states up to a principal quantum number $n_{\max} 
\sim 100$, resolving the angular momentum substates. This requires 
solving an $\mathcal{O}\left(10^4 \times 10^4\right)$ system of equations at each time step, which, even with modern computers, 
is extremely time consuming.

\section{New method of solution: the effective multi-level atom}
\label{section:new method}

We now give a computationally efficient method of solution for the primordial recombination problem. We factor the effect of the numerous transitions involving interior states in terms of effective transitions into and out of the much smaller number of interface states. Once the rates of these effective transitions are tabulated, the cosmological evolution of the free electron fraction can be obtained from a simple effective few-level atom calculation. We describe the method in Section~\ref{ss:general} and give the proof of its exact
equivalence to the standard MLA method in Section~\ref{sec:equivalence}.  In the subsequent Section~\ref{ss:nstar}, we consider which states should be treated as interface states.

\subsection{Motivations and general formulation}
\label{ss:general}

We first note that the only quantity of importance for CMB power spectrum calculations is the free electron fraction as a function of redshift, $x_e(z)$. The populations of the excited states are 
calculated only as an intermediate step -- if they are desired (e.g. to calculate H$\alpha$ scattering features \cite{HRS}), the populations of the excited states can be obtained by solving 
Eqs.~(\ref{eq:dot.X_K}, 
\ref{eq:dot.x_i}) once the free electron fraction is known. Furthermore, only the ``interface'' states $2s$ and $np$ are directly connected to the ground state and directly appear in the evolution equation for the free electron fraction Eq.~(\ref{eq:dot.x_e}). All other (``interior'') excited states are only connected with other excited states or with the continuum, through optically thin radiative transitions (and to a lesser extent through collisions \cite{CRMS07}). Interior states are only 
transitional states: an electron in the ``interior'' rapidly transitions through spontaneous and stimulated decays or absorptions caused by the blackbody radiation field (or collisions with free 
electrons and protons), until it is either photoionized, or reaches an interface state. There can be a very large number of transitions before any of these outcomes occurs, but the passage through the 
``interior'' is always very short compared to the overall recombination timescale, and can be considered as instantaneous (for the same reason that the steady-state approximation is valid in the standard MLA formulation).

Instead of computing the fraction of hydrogen in each interior state $K$, one can rather evaluate the probabilities that an atom initially in the interior state $K$ ultimately reaches one of the 
interface states or gets photoionized. Of course, after reaching an interface state, the atom may perfectly transition back to an interior state, or get photoionized. However, we consider the 
probability of \emph{first} reaching a given interface state before any other one, which is uniquely defined. For an atom in the interior state $K$, we denote by $P_K^i$ the probability of ultimately 
reaching the interface state $i$, and $P_{K}^e$ the probability of ultimately being photoionized. The probabilities $P_K^i$ must self-consistently account for both direct transitions $K \rightarrow i$ and all possible indirect transitions $K \rightarrow L \rightarrow i$ (with an arbitrary number of intermediate states). Mathematically, this translates to the system of linear equations:   
\beq
P_K^i = \sum_L \frac{R_{K \rightarrow L}}{\Gamma_K} P_L^i + \frac{R_{K \rightarrow i}}{ \Gamma_K },
\label{eq:PKi}
\eeq
where $\Gamma_K$ is the total width (or inverse lifetime) of the state $K$:
\beq
\Gamma_K \equiv \sum_L R_{K \rightarrow L} + \sum_j R_{K \rightarrow j} + \beta_K. \label{eq:GammaK}
\eeq
Similarly, the $P_K^e$ must satisfy the self-consistency relations:
\beq
P_K^e = \sum_L \frac{R_{K \rightarrow L}}{\Gamma_K} P_L^e + \frac{\beta_K}{ \Gamma_K },
\label{eq:PKe}
\eeq
We show in Appendix \ref{app:invertibility} that these linear systems are invertible and therefore uniquely determine $P^i_K$ and $P^e_K$. 
In Appendix~\ref{app:complementarity} we prove the complementarity 
relation,
\beq
\sum_i P_K^i + P_K^e = 1,
\label{eq:complementarity}
\eeq
which has the simple physical interpretation that an atom in the $K$th interior state eventually reaches an interface state or is photoionized with unit probability.

\begin{figure} 
\includegraphics[width = 86mm]{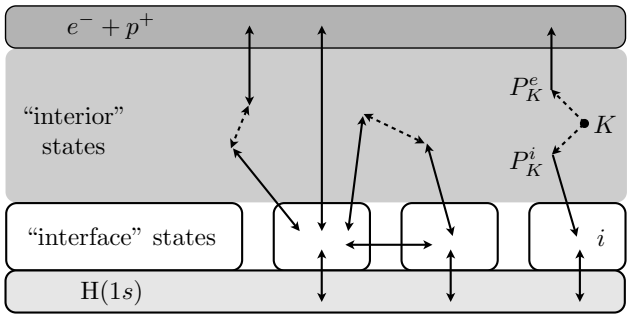}
\caption{Schematic representation of the formulation of the recombination problem adopted in this work. Dotted arrows represent possibly numerous fast transitions within the ``interior''.}
\label{fig:scheme}
\end{figure}

Once these probabilities are known, it is possible to describe the large number of transitions between all the states in a simplified manner, in terms of effective rates into and out of the interface states. To clarify the explanation, we illustrate in Figure \ref{fig:scheme} the processes described below.

An electron and a proton can effectively recombine to the interface state $i$ either through a direct recombination (with coefficient $\alpha_i$), or following a recombination to an interior state $K$ (with coefficient $\alpha_K$), from which a sequence of interior transitions may ultimately lead to the interface state $i$ with probability $P_K^i$. The effective recombination coefficient to the interface state $i$ is therefore:
\beq
\A_i \equiv \alpha_i + \sum_K \alpha_K P_K^i\label{eq:Ai}.
\eeq
Conversely, an atom in the interface state $i$ may effectively be ionized either through a direct photoionization (with rate $\beta_i$), or after being first excited to an interior state $K$ (with rate $R_{i \rightarrow K}$), from which the atom may ultimately be photoionized after a series of interior transitions with probability $P_K^e$. The effective photoionization rate from the interface state $i$ is therefore:
\beq
\B_i \equiv \beta_i + \sum_K R_{i\rightarrow K} P_K^e\label{eq:Bi}.
\eeq
Finally, atoms can effectively transition from an interface state $i$ to another interface state $j$, either through a direct transition if it is allowed, or after first transitioning through the interior. The effective transfer rate between the $i$th and $j$th interface states is therefore:
\beq
\R_{i\rightarrow j} \equiv R_{i \rightarrow j} + \sum_K R_{i \rightarrow K} P_K^j \ \ \ (j \neq i). \label{eq:Rij}
\eeq
The rate of change of the population of 
the interface state $i$ is therefore:
\barr
\dot x_i &=& x_e^2 n_{\rm H} \A_i + \sum_{j\neq i} x_j\R_{j \rightarrow i} + x_{1s} \tilde{R}_{1s \rightarrow i} \label{eq:dot.x_i.new}\\
&-& x_i \Bigl( \B_i + \sum_{j\neq i} \R_{i\rightarrow j} + \tilde{R}_{i\rightarrow 1s}\Bigr),\nonumber
\earr
where we have included the effective transitions described above, as well as transitions from and to the ground state.

The system of equations (\ref{eq:PKi}--\ref{eq:dot.x_i.new}) is exactly equivalent to the standard MLA formulation, as we shall show in 
Section ~\ref{sec:equivalence} below.

Let us now consider the dependences of the effective rates. In the purely radiative case, the probabilities $P_K^i$ and $P_K^e$ only depend on the 
radiation temperature $\Tr$, since transitions between excited states and photoionizations only depend on the locally thermal radiation field. As a 
consequence, the effective recombination rates $\mathcal{A}_i\left(\Tm, \Tr \right)$ are only functions of matter and radiation temperatures and the 
effective photoionization and bound-bound rates $\mathcal{B}_i\left(\Tr \right)$ and $\mathcal{R}_{i\rightarrow j}\left(\Tr \right)$ are functions of the 
radiation temperature only. When including collisional transitions, all effective rates become functions of the three variables $\Tr, \Tm$ and $n_e$. 
In all cases, effective rates can be easily tabulated and interpolated when needed for a recombination calculation.

Intuitively, we would expect that $\A_i$, $\B_i$, and $\R_{i\rightarrow j}$ satisfy the detailed balance relations,
\beq
g_i \rme^{-E_i/\Tr} \R_{i\rightarrow j}(\Tr) = g_j \rme^{-E_j/\Tr} \R_{j\rightarrow i}(\Tr)
\label{eq:DBR}
\eeq
and
\beq
g_i \rme^{-E_i/\Tr}\B_i(\Tr) = \frac{(2\pi\mu_e\Tr)^{3/2}}{h^3}  \A_i(\Tm=\Tr,\Tr).
\label{eq:DBAB}
\eeq
We show in Appendix~\ref{app:db} that these equations are indeed valid. This means that we only need to tabulate half of the 
$\R_{i\rightarrow j}$ [the other half can be obtained from Eq.~(\ref{eq:DBR})] and all the $\A_i$ [the $\B_i$ can be obtained from Eq.~(\ref{eq:DBAB}); in particular, we do not need to solve for the $P_K^e$]. 

We note that the probabilities $P_K^i, P_K^e$ are a generalization of the cascade matrix technique introduced by Seaton 
\cite{Seaton59}. Seaton's calculation assumed a vanishing ambient radiation field, so that electrons can only ``cascade down'' to lower energy 
states. In the context of the recombination of the primeval plasma, one cannot ignore the strong thermal radiation field, and electrons rather ``cascade 
up and down,'' following spontaneous and stimulated decays or photon absorption events. The spirit of our method is however identical to Seaton's 
cascade-capture equations \cite{Seaton59}, where the ``cascading'' process is decoupled from the particular process populating the excited states, or 
from the depopulation of the interface states.

\subsection{Equivalence with the standard MLA method} \label{sec:equivalence}

This section is dedicated to proving the equivalence of the EMLA equations, Eqs.~(\ref{eq:PKi}--\ref{eq:dot.x_i.new}), with the standard MLA 
equations, Eqs.~(\ref{eq:dot.X_K}, \ref{eq:dot.x_i}), in 
the steady-state limit for the interior states (i.e. where we set $\dot X_K\approx 0$). The steady-state approximation does not need to be made for the interface states to demonstrate the equivalence of the two formulations (but we do use it for practical computations since it is valid to very high accuracy).

We denote by $N$ the number of interior states and $n_*$ the number of interface states (we will address in Section \ref{ss:nstar} the issue of which states need to be considered as interface states).

We begin by defining the $N\times N$ rate matrix ${\bf M}$ whose elements are
\beq
M_{KL} \equiv \delta_{KL}\Gamma_K - \left(1 - \delta_{KL}\right) R_{K \rightarrow L}.
\label{eq:MKL}
\eeq
We also define the $n_* + 1$ length-$N$ vectors ${\bf P}^i, {\bf P}^e$ whose elements are the probabilities  $P_K^i$ and $P_K^e$ respectively, and the $n_*+1$ length-$N$ vectors ${\bf R}^i, {\bf R}^e$ of components
\barr
R^i_K &\equiv& R_{K\rightarrow i},\\
R_K^e &\equiv& \beta_K.
\earr
The defining equations for the probabilities, Eqs.~(\ref{eq:PKi}, \ref{eq:PKe}), can be written in matrix form ${\bf MP}^i={\bf R}^i$ and $\bf{MP}^e = {\bf R}^e$ respectively (after multiplication by $\Gamma_K$). We show in Appendix \ref{app:invertibility} that the matrix ${\bf M}(T_{\rm r})$ is invertible, for any temperature $\Tr \geq 0$. The formal solutions for the probabilities are therefore
\barr
{\bf P}^i &=& {\bf M}^{-1}{\bf R}^i \label{eq:P-solve}\\
{\bf P}^e &=& {\bf M}^{-1}{\bf R}^e.\label{eq:Pe-solve}
\earr
We also define the length-$N$ vector ${\bf X}$ which contains the populations of the interior states $X_K$, and the length-$N$ vector ${\bf S}$ of components
\beq
S_K \equiv x_e^2 n_{\rm H} \alpha_K + \sum_j x_j R_{j \rightarrow K}. \label{eq:S_K}
\eeq
A careful look at Eq.~(\ref{eq:dot.X_K}) in the steady-state
approximation ($\dot X_K=0$) shows that it is the matrix equation ${\bf M}^{\rm T}{\bf X} = {\bf S}$, which has the solution:
\beq
{\bf X}=\left({\bf M}^{\rm T}\right)^{-1}{\bf S} = \left({\bf M}^{-1}\right)^{\rm T}{\bf S}.
\label{eq:X-solve}
\eeq
Both Eqs.~(\ref{eq:dot.x_i}) and (\ref{eq:dot.x_i.new}) can be cast in the form
\barr
\dot{x}_i &=& x_e^2 n_{\rm H} \alpha_i + \sum_{j\neq i} x_i R_{j\rightarrow i} + x_{1s} \tilde{R}_{1s\rightarrow i} \nonumber\\
      &-& x_i \Big{(}\beta_i + \sum_{j\neq i} R_{i\rightarrow j} + \tilde{R}_{i \rightarrow 1s}\Big{)} + \dot{x}_i|_{\rm interior}.
\earr
The only a priori different term is the net transition rate from the interior to the state $i$, $\dot{x}_i|_{\rm interior}$. In the standard MLA formulation, Eq.~(\ref{eq:dot.x_i}), this term is
\barr
\dot{x}_i|_{\rm interior}^{(\rm MLA)} &=& \sum_K \left(X_K R_K^i - x_i R_{i\rightarrow K}\right)\\
&=& {\bf X}^{\rm T} {\bf R}^i - x_i \sum_K R_{i \rightarrow K}. \label{eq:xdot.interior.mla}
\earr
With our new formulation, Eq.~(\ref{eq:dot.x_i.new}), using the definitions of the effective rates Eqs.~(\ref{eq:Ai}--\ref{eq:Rij}), the net transition rate from the interior to the state $i$ is:
\barr
\dot{x}_i|_{\rm interior}^{(\rm EMLA)} &=& \sum_K \Bigg{[}x_e^2 n_{\rm H}\alpha_K P_K^i + \sum_{j\neq i}x_j R_{j\rightarrow K}P_K^i\nonumber\\
&-& x_i R_{i\rightarrow K} (P_K^e + \sum_{j \neq i} P_K^j)\Bigg{]}.\label{eq:dot.x_i.interior.new}
\earr
Using the complementarity relation Eq.~(\ref{eq:complementarity}), we rewrite $P_K^e + \sum_{j \neq i} P_K^j = 1 - P_K^i$. We then recognize that the common factor of $P_K^i$ is just the $K$-th component of the vector ${\bf S}$, Eq.~(\ref{eq:S_K}), so we can rewrite Eq.~(\ref{eq:dot.x_i.interior.new}) as
\barr
\dot{x}_i|_{\rm interior}^{(\rm EMLA)} = {\bf S}^{\rm T} {\bf P}^i - x_i\sum_K R_{i\rightarrow K}. \label{eq:xdot.interior.fla}
\earr
From the formal solution for the populations of the interior states Eq.~(\ref{eq:X-solve}), we have
\beq
{\bf X}^{\rm T} {\bf R}^i = {\bf S}^{\rm T}{\bf M}^{-1} {\bf R}^i = {\bf S}^{\rm T} {\bf P}^i,
\eeq
where the second equality is obtained from the formal solution for the probabilities $P_K^i$, Eq.~(\ref{eq:P-solve}). We therefore see from Eqs.~(\ref{eq:xdot.interior.mla}) and (\ref{eq:xdot.interior.fla}) that 
\beq
\dot{x}_i|_{\rm interior}^{(\rm MLA)} = \dot{x}_i|_{\rm interior}^{(\rm EMLA)}, 
\eeq
and hence the two formulations are \emph{exactly} equivalent. They only differ by the order in which the bilinear product $ {\bf S}^{\rm T}{\bf M}^{-1} {\bf R}^i $ is evaluated.

\subsection{Choice of interface states}
\label{ss:nstar}

If one naively includes all $np$ states up to $n = n_{\max} =\mathcal{O}(100)$ in the list of interface states, the interpolation of effective rates can become somewhat cumbersome as it involves 
$\mathcal{O}\left(10^4\right)$ functions of one to three variables.  However, only the lowest few of these states actually have significant transition rates to the ground state; indeed, most of the 
decays to the ground state proceed through either $2s$ (two-photon decay) or $2p$ (Ly$\alpha$ escape), as anticipated in the earliest studies \cite{Peebles, Zeldovich_et_al}.

The rate of Lyman line escape is dominated by the lowest few lines.  For example, if the relative populations of the excited states were given by the Boltzmann ratios (which is a good approximation until late times) then the net decay rate in the $np \rightarrow 1s$ transition (not accounting for feedback from the next line) would be proportional to
\beq
\dot x_{np\rightarrow 1s} \propto (1-n^{-2})^3 e^{-E_n/\Tr}.
\label{eq:Boltzmann}
\eeq
This relation would imply that the Ly$\beta$ escape rate is $<1$\% of the Ly$\alpha$ escape rate, and the higher-order Lyman lines contribute even less.  Our previous computations of the escape rates (e.g. 
Ref.~\cite{Hirata_2photon}) agree with this expectation. These considerations imply that for $n \geq 3,~ \dot{x}_{1s}|_{np} \ll \dot{x}_{1s}|_{2p}$ in Eq.~(\ref{eq:dot.x_e}). Moreover, an atom in the $np$ state with $n \geq 3$ is much more likely to spontaneously decay to $n's$ or $n'd$, with $2 \leq n' < n$, than to emit a Lyman-$n$ photon that successfully escapes the line. This implies that $\Big{|}\dot{x}_{np}|_{1s}\Big{|} \ll \Big{|}\dot{x}_{np}\Big{|}$ in Eq.~(\ref{eq:dot.x_i}).
  
In addition to a very low net decay rate out of the $np$ states for $n \geq 3$, feedback between neighboring lines further suppresses their efficiency as interface states. The few photons that escape the 
Ly$(n+1)$ line will be reabsorbed almost certainly in the next lower line, after a redshift interval
\beq
\Delta z = z_{\rm em} - z_{\rm ab} = (1 + z_{\rm ab}) \left(\frac{E_{n+1,1}}{E_{n1}} - 1\right).
\eeq 
Feedback between the lowest-lying lines is not instantaneous: $\Delta z/ (1 + z_{\rm ab}) = 0.185$ for Ly$\beta \rightarrow$Ly$\alpha$ feedback, $0.055$ for Ly$\gamma \rightarrow$Ly$\beta$, and 
0.024 for Ly$\delta \rightarrow$Ly$\gamma$. However, for higher-order lines, feedback rapidly becomes nearly instantaneous as $\Delta z /(1 + z_{\rm ab}) \sim 2/n^3$.
Thus the effect of the higher Lyman lines is even weaker than Eq.~(\ref{eq:Boltzmann}) would suggest.
Recent work \cite{Kholupenko_Deuterium} has shown that including lines above Ly$\beta$ results in a fractional error $|\Delta x_e|/x_e$ of at most $\approx 3\times 10^{-4}$.

We therefore conclude that very accurate recombination histories can be obtained by 
only including $2s, 2p ,..., n_*p$ as interface states and negelecting higher-order Lyman transitions altogether.
We will use $n_*=3$ in this paper, and investigate the optimal value of this cutoff more quantitively in future work.

Our formulation in terms of effective transition rates and interface states is therefore much better adapted for a fast recombination calculation that 
the standard MLA formulation. To compute accurate recombination histories, explicitly accounting for high-$n$ shells of hydrogen, one first needs to 
tabulate the $\{\A_i\}$ and $\{\R_{i\rightarrow j}\}$ on temperature grids.
The computation of the effective rates is the time-consuming 
part of the calculation; however, since they are independent of the cosmological parameters, this can be done once, and not repeated for each cosmology.  The free electron fraction can 
then be computed very quickly for any given cosmology by solving the $n_* + 1$ equations (\ref{eq:dot.x_i.new}) and (\ref{eq:dot.x_e}), interpolating 
the effective rates from the precomputed tables. Note that Eq.~(\ref{eq:dot.x_i.new}) is a simple $n_* \times n_*$ system of linear algebraic equations 
in the steady-state approximation.

\section{Implementation and results}
\label{section:results}
Here we give some details on the implementation of our EMLA code. Section \ref{sec:eff.rates.comp} describes the computation of the effective rates (the computationally expensive part of the calculation, which needs to be done only once). Section \ref{sec:efla.code} descibes the implementation of the ultrafast effective few-level atom calculation. We show our recombination histories and compare our results with the existing standard MLA code \textsc{RecSparse} \cite{Grin_Hirata} in Section \ref{sec:compare}.

\subsection{Computation of the effective rates}\label{sec:eff.rates.comp}

We have implemented the calculation of the effective rates in the purely radiative case. Bound-free rates were computed by numerically integrating 
Eq.~(\ref{eq:alpha}) using an 11-point Newton-Cotes method, where the radial matrix elements $g(n, l, \kappa, l')$ were obtained using the recursion 
relation given by Burgess \cite{Burgess}. Einstein $A$-coefficients were computed by using the recursion relations obtained by Hey \cite{Hey06} for the 
radial matrix elements $R_{n'l'}^{nl}$. Finally, we obtained the probabilities $P_K^i$ using a sparse matrix technique similar to that of 
Ref.~\cite{Grin_Hirata} when solving Eq.~(\ref{eq:PKi}). We accounted explicitly for all excited states up to a principal quantum number $n_{\max}$, resolving angular momentum substates. We tabulated the effective rates ${\cal A}_i(\Tm,\Tr)$ on a grid of 200 log-spaced points in $\Tr$ from 0.04 to 0.5 eV and 20 linearly spaced points 
in $\Tm/\Tr$ from 0.8 to 1.0, and ${\cal R}_{i\rightarrow j}(\Tr)$ on the grid of points in $\Tr$. The maximum 
relative change in the effective rates ($\Delta\ln\A_i$ or $\Delta\ln\R_{i\rightarrow j}$) over the whole range of temperatures considered is 
0.051 when comparing $n_{\max}=64$ vs. 128, 0.015 when comparing $n_{\rm max}=128$ vs. 250, and 0.005 when comparing $n_{\rm max}=250$ vs. 500.

\begin{figure*} 
\includegraphics[width = 86mm]{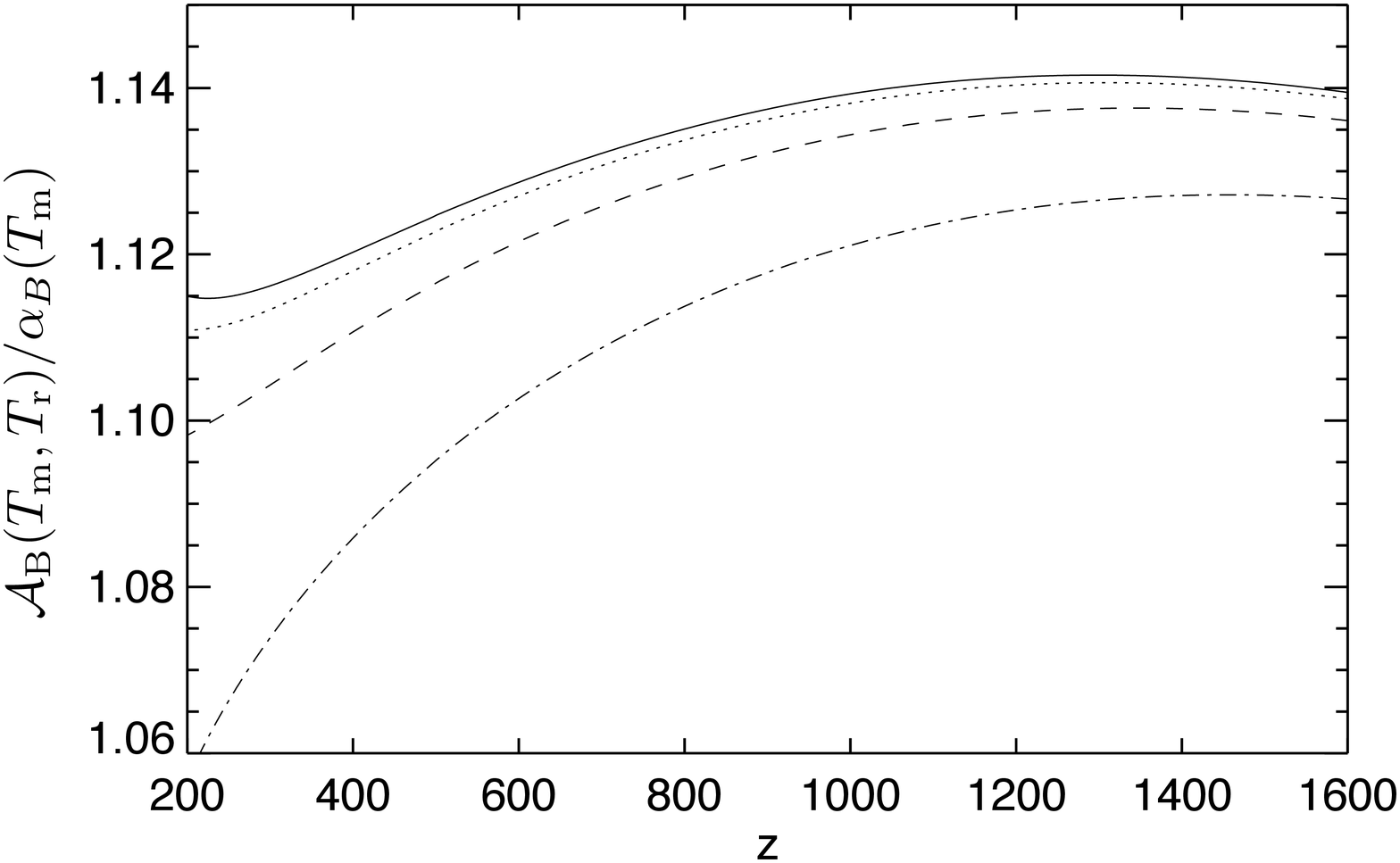}
\includegraphics[width = 86mm]{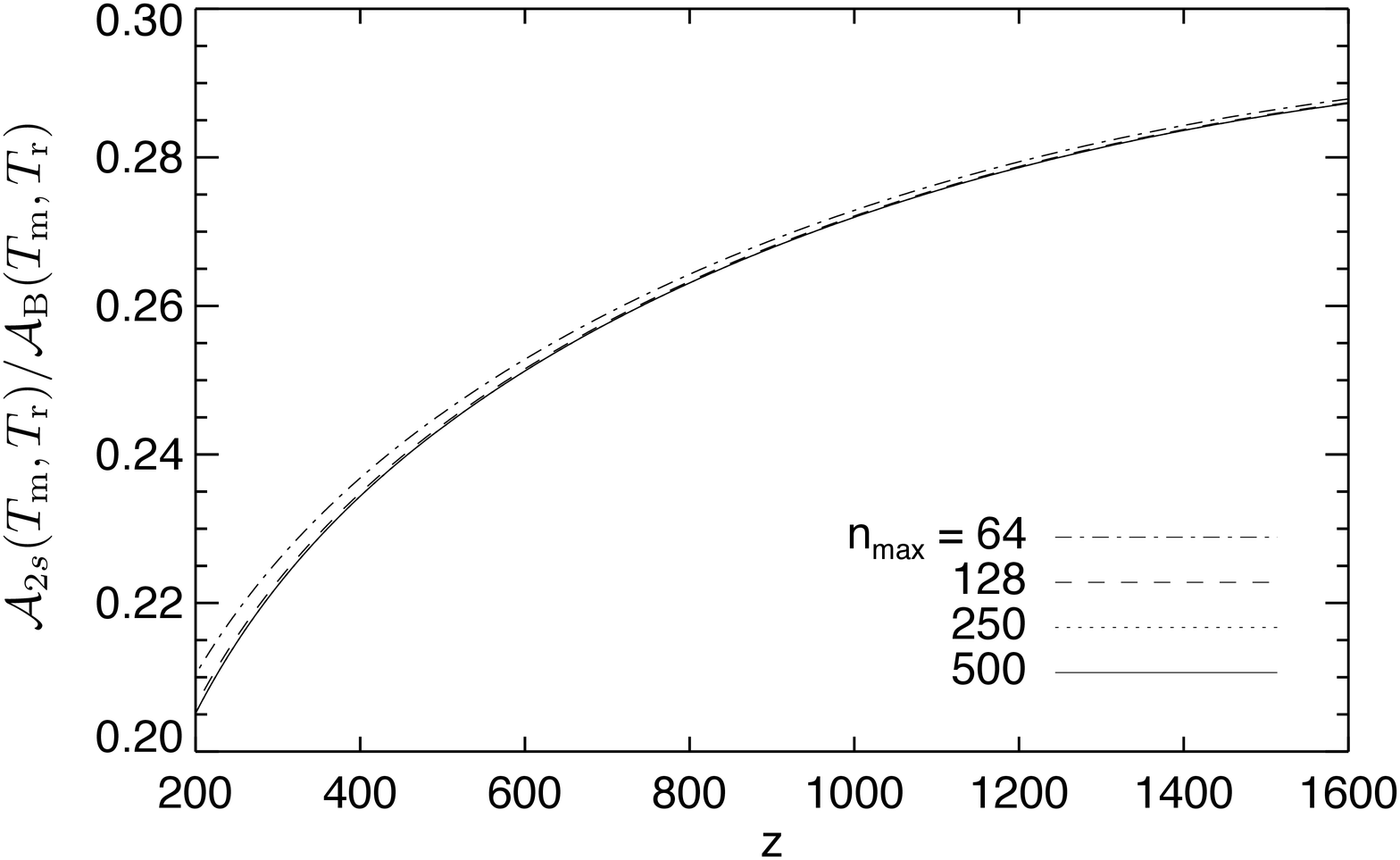}
\caption{\emph{Left panel}: ``Exact fudge factor'' as a function of redshift $\mathcal{A}_{\rm B}(\Tm, \Tr) / \alpha_{\rm B}(\Tm)$, for several values of $n_{\max}$, using $\Tm(z)$ computed by \textsc{RecSparse} for cosmological parameters as 
in Ref.~\cite{Grin_Hirata}. We use the fit of Ref.~\cite{alphaB} for the case-B recombination coefficient $\alpha_{\rm B}(\Tm)$. For 
comparison, the code \textsc{Recfast} uses a constant fudge factor $F = 1.14$ to mimick the effect of high-$n$ shells. \emph{Right panel}: Fraction of the effective recombinations to the $n=2$ shell that lead to atomic hydrogen in the $2s$ state. In both cases the effective rates were computed for $n_* = 2$, i.e. with interface states $2s$ and $2p$ only, neglecting escape from the Lyman $\beta,\gamma,...$ lines.}
\label{fig:fudge}
\end{figure*}

In the left panel of Figure \ref{fig:fudge}, we show the total effective recombination coefficient $\mathcal{A}_{\rm B}(\Tm, \Tr) \equiv \mathcal{A}_{2s}(\Tm, \Tr) + \mathcal{A}_{2p}(\Tm, \Tr)$ computed for $n_* = 2$ (i.e. with interfaces states $2s$ and $2p$ only, neglecting all Lyman transitions above Lyman-$\alpha$), normalized to the case-B recombination coefficient $\alpha_{\rm B}(\Tm)$. Note that $\alpha_{\rm B}(\Tm)$ is just $\mathcal{A}_{\rm B}(\Tm, \Tr = 0| n_{\max} = \infty)$ with our notation; indeed, for $\Tr = 0$, $\beta_K = 0$ and therefore $P_K^e = 0$ for all $K$ so $\sum_i P_K^i = 1$ and hence $\sum_i\mathcal{A}_i = \sum_i \alpha_i + \sum_K \alpha_K = \sum_{nl} \alpha_{nl}$, where the last sum is over all excited states. We can see that as the radiation temperature increases (i.e. as the redshift increases in Figure \ref{fig:fudge}), the convergence with $n_{\max}$ becomes faster. This is to be expected, since for higher $\Tr$, highly excited hydrogen is more easily photoionized, i.e. $P_{nl}^e$ becomes closer to unity. In that case adding more shells to the calculation does not matter so much because recombinations to the highest shells are very inefficient, due to the high probability of a subsequent photoionization.

In the right panel of Figure \ref{fig:fudge}, we show the ratio $\mathcal{A}_{2s}(\Tm, \Tr)/\mathcal{A}_{\rm B}(\Tm, \Tr)$, which is the fraction of recombinations to the $n = 2$ shell that are to the $2s$ level. This fraction is in general different from the intuitive value of 1/4, and its exact value depends on temperature.

\subsection{Ultrafast EMLA code}\label{sec:efla.code}

In order to actually compute the recombination history, we require an evolution equation for the free electron fraction,
\begin{equation}
\dot{x}_e(x_e, n_{\rm H}, H, \Tm, \Tr),
\end{equation}
and in some cases a similar equation for $\dot{T}_{\rm m}$.  For concreteness, we implement the case of 3 interface states $i\in\{2s,2p,3p\}$ ($n_*=3$).

To compute $\dot x_e$, we first obtain the downward ${\cal R}_{i\rightarrow j}(\Tr)$ from our table via cubic 
polynomial (4-point) interpolation and ${\cal A}_{i}(\Tm,\Tr)$ via bicubic interpolation (2-dimensional in $\ln\Tr$ and $\Tm/\Tr$ using $4\times 4$ 
points).  The upward ${\cal R}_{j\rightarrow i}(\Tr)$ are obtained using Eq.~(\ref{eq:DBR}) and the effective photoionization rates ${\cal B}_i(\Tr)$ are obtained using Eq.~(\ref{eq:DBAB}).  We then solve for the $\{x_i\}$ using 
Eq.~(\ref{eq:dot.x_i.new}), and finally obtain $\dot x_e$ using Eq.~(\ref{eq:dot.x_e}).

The matter temperature is determined by the Compton evolution 
equation,
\beq
\dot{T}_{\rm m} = -2H\Tm + \frac{8\sigma_{\rm T}a_{\rm r}\Tr^4x_e(\Tr-\Tm)}{3(1+f_{\rm He}+x_e)m_ec},
\label{eq:Tm-evol}
\eeq
where $\sigma_{\rm T}$ is the Thomson cross section, $a_{\rm r}$ is the radiation constant, $f_{\rm He}$ is the He:H ratio by number of nuclei, $m_e$ 
is the electron mass, and $c$ is the speed of light.  At high redshift, one may use the steady-state solution (see Appendix A of 
Ref.~\cite{Hirata_2photon}),
\beq
\Tm \approx T_{\rm m,ss} = \Tr \left[ 1 + \frac{3(1+f_{\rm He}+x_e)m_ec H}{8\sigma_{\rm T}a_{\rm r}\Tr^4x_e} \right]^{-1}.
\label{eq:Tm-ss}
\eeq

At the highest redshifts, the ODE describing hydrogen recombination is stiff; therefore for $z > 1570$ we follow the recombination history using perturbation theory around the Saha approximation, as described in Appendix \ref{appendix:post-saha}.  At $500<z<1570$ we use Eq.~(\ref{eq:Tm-ss}) to set the matter temperature, and a 
fourth-order Runge-Kutta integration algorithm (RK4) to follow the single ODE for $x_e(z)$; and at $z<500$ we use RK4 to follow the two ODEs for 
$x_e(z)$ and $\Tm(z)$ simultaneously.  The integration step size is $\Delta z = -1.0$ (negative since we go from high to low redshifts).

\subsection{Results and code comparison} \label{sec:compare}

\begin{figure*}
\includegraphics[width=86mm]{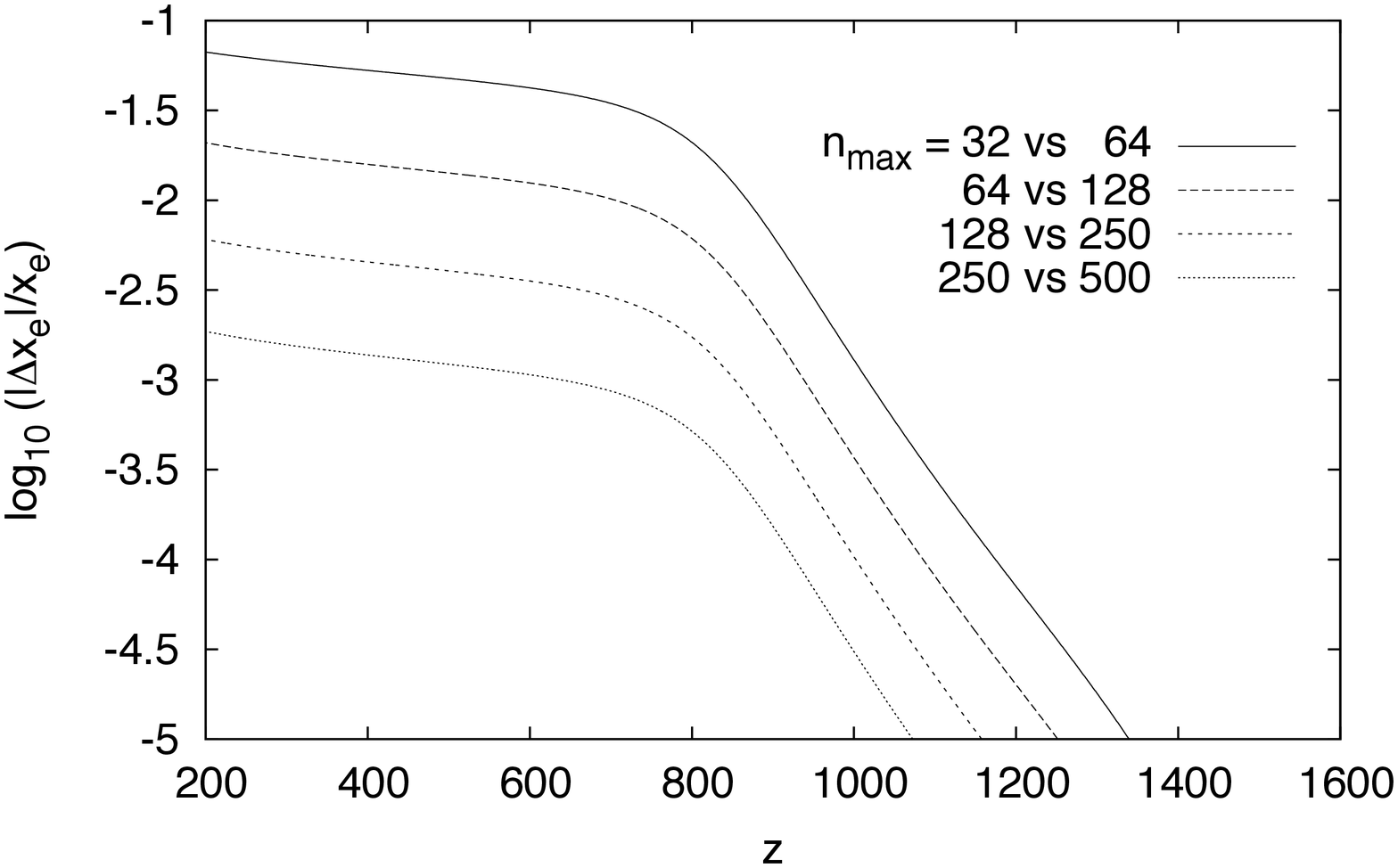}
\includegraphics[width=86mm]{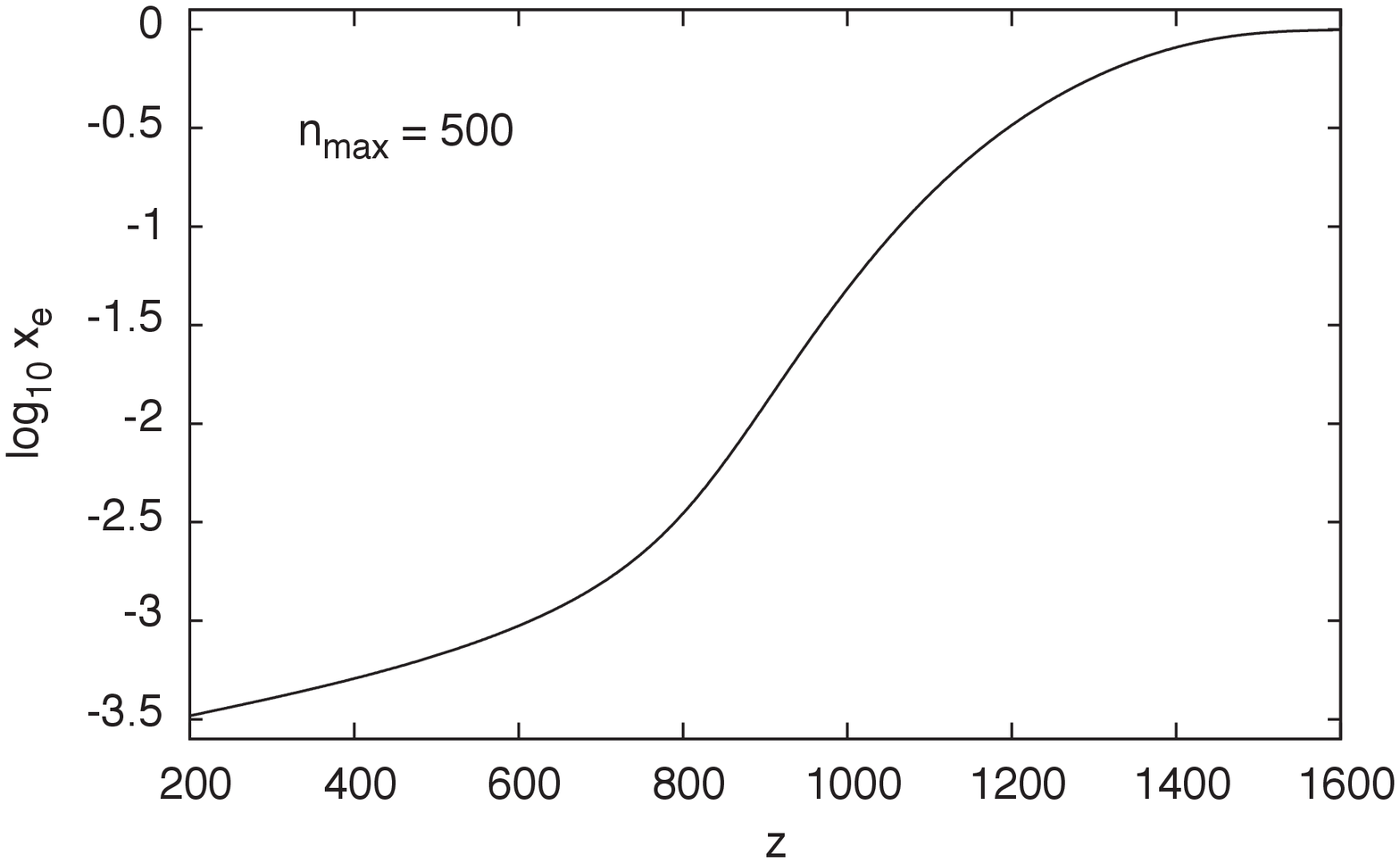}
\caption{\label{fig:xehist}\emph{Left panel}: Relative differences between recombination histories computed with successively more accurate effective rates. \emph{Right panel}: Recombination history for effective rates computed with $n_{\max} = 500$, i.e. accounting explicitly for 125,250 states of the hydrogen atom.}
\end{figure*}

\begin{figure}
\includegraphics[width=86mm]{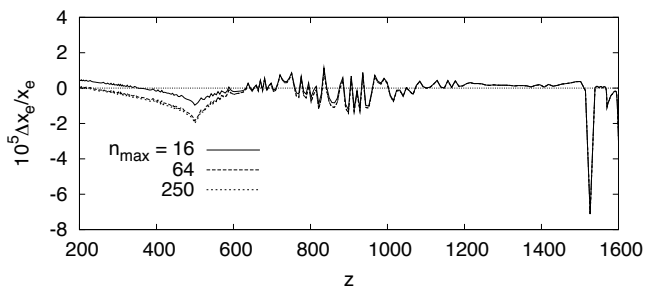}
\caption{\label{fig:ufcompare}A comparison of our ultrafast code to \textsc{RecSparse} \cite{Grin_Hirata}, for different values of $n_{\max}$.  The vertical axis is the fractional
difference in free electron abundance rescaled by $10^5$ (positive indicating that \textsc{RecSparse} gives a larger $x_e$). We see that the maximum fractional
deviation is $<8\times 10^{-5}$.  The feature around $z = 1540$ is due to a timestep change in \textsc{RecSparse}.}
\end{figure}

We have tabulated the effective rates for $n_{\max} = 16$, 32, 64, 128, 250 and 500. It is in principle possible to compute the effective rates for an arbitrarily high $n_{\max}$, but it is not meaningful to do so as long as collisional transitions are not properly accounted for. The recurring computation time of our ultrafast EMLA code is 0.08 seconds on a MacBook laptop computer with a 2.1 GHz processor, independently of $n_{\max}$.  Our recombination histories are shown in Figure~\ref{fig:xehist}. We compared our results with the existing standard MLA code \textsc{RecSparse} for $n_{\max} = 16$, 32, 64, 128 and 250. As can be seen in Figure~\ref{fig:ufcompare}, the two codes agree to better than $8\times 10^{-5}$
across the range $200<z<1600$, despite having different methods for accounting for the excited states, and independent implementations for matrix elements and ODE integration.

\section{Conclusions and future directions}
\label{section:conclusion}

We have shown that the computation of primordial hydrogen recombination can be factored into two independent calculations. On the one hand, most 
excited states are not directly radiatively connected to the ground state, and undergo transitions caused by the thermal bath of blackbody photons at 
the relevant frequencies, as well as the thermal electrons and protons. One can account for these numerous transitions with effective transition rates into and out of the ``interface'' 
states which are connected to the ground state. The computationally intensive aspect 
of a recombination calculation in fact resides in the evaluation of these effective rates, which are functions of matter and radiation temperature only. This calculation being independent of cosmological 
parameters, it can be done prior to any recombination calculation, once and for all. A simple effective few-level atom can then be evolved for any set of 
cosmological parameters, without any need for ``fudge factors'' or approximations.

This work does \emph{not} present a final recombination code satisfying the accuracy requirements for future CMB experiments. Firstly, collisional 
transitions were not included. They may be particularly important for the high-$n$ states. The effective rates computed here are therefore only 
approximating the correct rates in the limit of zero density. Our formalism is general and collisions can be included as soon as accurate rates are 
available (the main change would be that the interpolation tables would require $\ln n_e$ as an additional independent variable).  Secondly, we have not included important radiative transfer effects, 
such as feedback between low-lying Lyman lines \cite{CS07, 
Kholupenko_Deuterium}, two-photon decays from $n\ge 3$ \cite{DG05, WS07, CS08, Hirata_2photon, CS09b}, resonant scattering in 
Ly$\alpha$ \cite{GD08, Hirata_Forbes, CS09c}, or overlap of the high-lying Lyman lines (work in preparation). To preserve the computational efficiency of 
our method, fast analytic approximations have to be developed to include these effects, which will be the subject of future work.

\section*{Acknowledgements}

We thank Dan Grin for numerous useful and stimulating conversations, and for providing data from \textsc{RecSparse} computations for code comparison. We 
also aknowledge fruitful conversations with the participants of the July 2009 Paris Workshop on Cosmological Recombination. We thank Dan Grin, Marc Kamionkowski and Jens Chluba for a careful reading of the draft of this paper. Y. A.-H. and C.H. are 
supported by the U.S. Department of Energy (DE-FG03-92-ER40701) and the National Science Foundation (AST-0807337). C. H. is supported by the Alfred P. 
Sloan Foundation.

\appendix

\section{Invertibility of the system defining the $P_K^i, P_K^e$.}\label{app:invertibility}

In this section we show that the matrix ${\bf M}(T_{\rm r})$ defined in Eq.~(\ref{eq:MKL}) is non-singular, for any value of the radiation temperature $T_{\rm r} \geq 0$.

Let us consider the eigenvalue equation ${\bf Mb} = {\bf 0}$ and select a particular $K_1$ such that $|b_{K_1}| \geq |b_L|$ for all $L$. The eigenvalue equation implies
\barr
0 &=& \big{|} M_{K_1K_1} b_{K_1} + \sum_{L \neq K} M_{K_1L}b_L\big{|} \nonumber \\
  &\geq& M_{K_1 K_1} \big{|} b_{K_1} \big{|} - \sum_{L \neq K_1}\big{|}M_{K_1 L}\big{|} \big{|} b_L\big{|}, \label{eq:zero.gt.ineq}
\earr 
where we have used the inverse triangle inequality. The matrix ${\bf M}$ is diagonally dominant, i.e. 
\beq
\forall K, ~~ M_{KK} = \Gamma_K \geq \sum_{L\neq K}R_{K\rightarrow L} = \sum_{L\neq K} |M_{KL}| \label{eq:M.diag.dominant}
\eeq
Using the inequality (\ref{eq:M.diag.dominant}) for $K = K_1$ in Eq.~(\ref{eq:zero.gt.ineq}), we obtain
\beq
0 \geq \sum_{L \neq K_1}\big{|}M_{K_1 L}\big{|} \left(|b_{K_1}| - |b_L|\right).\label{eq:K_1}
\eeq
For any interior state $K_1$, there always exists a sequence of transitions that ultimately leads to some interface state $i$, $K_1 \rightarrow K_2 \rightarrow ... \rightarrow K_n \rightarrow i$, for any temperature $\Tr \geq 0$. (i.e. there are no ``dead end'' interior states). In particular, $\big{|}M_{K_1 K_2}\big{|} = R_{K_1 \rightarrow K_2} > 0$. For Eq.~(\ref{eq:K_1}) to hold, it is therefore necessary that $|b_{K_1}| = |b_{K_2}|$. Repeating the above reasoning recursively leads to $|b_{K_1}| = |b_{K_2}| = ... = |b_{K_n}|$.

For the last interior state of this sequence, $K_n$, the inequality (\ref{eq:M.diag.dominant}) is strict since $R_{K_n \rightarrow i} > 0$. The eigenvalue equation projected on $K_n$ leads to Eq.~(\ref{eq:zero.gt.ineq}) for $K_n$:
\beq
0 \geq  M_{K_n K_n} \big{|} b_{K_n} \big{|} - \sum_{L \neq K_n}\big{|}M_{K_n L}\big{|} \big{|} b_L\big{|}.\label{eq:K_n}
\eeq
If ${\bf b} \neq {\bf 0}$, then $|b_{K_n}| > 0$ and the strict inequality (\ref{eq:M.diag.dominant}) for $K = K_n$ used in Eq.~(\ref{eq:K_n}) implies the contradictory result
\beq
0 > \sum_{L \neq K_n}\big{|}M_{K_n L}\big{|} \left(|b_{K_n}| - |b_L|\right) \geq 0. 
\eeq  
As a consequence, ${\bf M b} = {\bf 0}$ implies that ${\bf b} = {\bf 0}$ necessarily. This proves that ${\bf M}(\Tr) $ is nonsingular, for any $\Tr \geq 0$.

\section{Proof of the complementarity relation $\sum_i P_K^i + P_K^e = 1$} \label{app:complementarity}

We define the length-$N$ vector ${\bf V}\equiv(1,1,...,1)^{\rm T}$, and note that
\beq
({\bf MV})_K = \sum_L M_{KL} = \sum_j R_{K\rightarrow j} + \beta_K
\eeq
(the $R_{K\rightarrow L}$ terms cancel). In matrix form, this reads:
\barr
{\bf MV} = \sum_j {\bf R}^j + {\bf R}^e = {\bf M}\left[\sum_j {\bf P}^j + {\bf P}^e\right].
\earr
The matrix ${\bf M}$ being invertible, this implies $\sum_j {\bf P}^j + {\bf P}^e = {\bf V}$, which once projected on each component $K$ is just the complementarity relation Eq.~(\ref{eq:complementarity}).

\section{Detailed balance relations}
\label{app:db}

This appendix is dedicated to proving the detailed balance relations for ${\cal R}_{i\rightarrow j}$ and ${\cal B}_i$. Defining the contributions of individual states to the partition function,
\beq
Q_K \equiv g_K e^{-E_K/\Tr}
\eeq
and similarly for $Q_i$, we make use of the standard principle of detailed balance for rates connecting individual states,
\beq
Q_K R_{K\rightarrow L} = Q_L R_{L\rightarrow K},
\label{eq:QR}
\eeq
and similarly $Q_KR_{K\rightarrow i}=Q_iR_{i\rightarrow K}$.

We begin by defining the $N\times N$ nonsingular diagonal matrix ${\bf F}$ that is proportional to the equilibrium abundances,
\beq
F_{KL} \equiv Q_K\delta_{KL}.
\eeq
Then Eq.~(\ref{eq:QR})
combined with the definition Eq.~(\ref{eq:MKL}) implies that ${\bf FM}$ is {\em symmetric}.  It therefore follows that its matrix inverse ${\bf 
M}^{-1}{\bf F}^{-1}$ is symmetric, and hence that
\beq
\frac{({\bf M}^{-1})_{KL}}{Q_L} = \frac{({\bf M}^{-1})_{LK}}{Q_K}.
\label{eq:temp2}
\eeq

The transition rate, Eq.~(\ref{eq:Rij}), can be expanded using Eq.~(\ref{eq:P-solve}) as
\beq
{\cal R}_{i\rightarrow j} = R_{i\rightarrow j} + \sum_{K,L} ({\bf M}^{-1})_{KL} R_{i\rightarrow K} R_{L\rightarrow j}.
\eeq
We then see that:
\barr
Q_i{\cal R}_{i\rightarrow j} &=& Q_iR_{i\rightarrow j} + \sum_{K,L} Q_i ({\bf M}^{-1})_{KL} R_{i\rightarrow K} R_{L\rightarrow j}
\nonumber \\
&=& Q_iR_{i\rightarrow j} + \sum_{K,L} Q_K ({\bf M}^{-1})_{KL} R_{K\rightarrow i} R_{L\rightarrow j}
\nonumber \\
&=& Q_jR_{j\rightarrow i} + \sum_{K,L} Q_L ({\bf M}^{-1})_{LK} R_{K\rightarrow i} R_{L\rightarrow j}
\nonumber \\
&=& Q_jR_{j\rightarrow i} + \sum_{K,L} Q_j ({\bf M}^{-1})_{LK} R_{K\rightarrow i} R_{j\rightarrow L}
\nonumber \\
&=& Q_j{\cal R}_{j\rightarrow i},
\earr
where we have used Eq.~(\ref{eq:QR}) twice and in the third equality used Eq.~(\ref{eq:temp2}).
This proves Eq.~(\ref{eq:DBR}).

We may also relate the effective recombination and photoionization rates.  To do so, we consider the case of $\Tm=\Tr$ and define
\beq
q \equiv \left( \frac{ 2\pi\mu_e\Tr }{ h^2 }\right)^{3/2}.
\eeq
Then Eq.~(\ref{eq:beta-db}) can be written as $Q_{nl}\beta_{nl} = q\alpha_{nl}$.  Using Eq.~(\ref{eq:Pe-solve}), we see that
\barr
Q_i \B_i &=& Q_i\beta_i+\sum_{K,L} Q_i R_{i\rightarrow K} ({\bf M}^{-1})_{KL}\beta_L
\nonumber \\
&=& q\alpha_i + \sum_{K,L} Q_K R_{K\rightarrow i} ({\bf M}^{-1})_{KL}\beta_L
\nonumber \\
&=& q\alpha_i + \sum_{K,L} Q_L R_{K\rightarrow i} ({\bf M}^{-1})_{LK}\beta_L
\nonumber \\
&=& q\alpha_i + \sum_{K,L} q R_{K\rightarrow i} ({\bf M}^{-1})_{LK}\alpha_L
\nonumber \\
&=& q\A_i,
\earr
where in the last equality we have used Eq.~(\ref{eq:Ai}) with the $P^i_K$ determined by Eq.~(\ref{eq:P-solve}).  This proves Eq.~(\ref{eq:DBAB}).

\section{Implementation of post-Saha correction at early times}\label{appendix:post-saha}
At the highest redshifts, the ODE describing hydrogen recombination is stiff, and we follow the recombination history using perturbation theory around the Saha approximation, which we describe here. The idea is that the actual ionization fraction $x_e(z)$ is slightly greater than the Saha equation would predict because there are more recombinations than ionizations ($\dot x_e<0$) and a slight deviation from thermodynamic equilibrium is required in order to drive this imbalance.  We may thus take an ODE for
the recombination history (for simplicity we use the Peebles ODE \cite{Peebles}
with an updated recombination coefficient $\alpha_{\rm B}$ \cite{alphaB}),
and Taylor-expand it around the Saha ionization fraction:
\beq
\dot x_e^{\rm P}(x_e, z) = D_1(x_e-x_e^{\rm Saha}) +{\cal O}(x_e-x_e^{\rm Saha})^2.
\label{eq:xepdot}
\eeq
Here the superscript ``$^{\rm P}$'' denotes the Peebles ODE, and the zeroeth-order coefficient in the Taylor series vanishes since thermal equilibrium considerations imply $\dot x_e^{\rm P}(x_e^{\rm Saha}, z) = 0$.  The coefficient $D_1$ may be obtained by numerical differentiation; we use a two-sided finite difference with $\Delta x_e=\pm0.01x_e^{\rm Saha}(1-x_e^{\rm Saha})$.  Then we obtain a post-Saha corrected solution by setting the left-hand side of Eq.~(\ref{eq:xepdot}) with $\dot x_e^{\rm Saha}(z)$ (again obtained by a two-sided finite difference with $\Delta z=\pm1$):
\beq
x_e^{\rm corr}(z) = x_e^{\rm Saha}(z) + \frac{\dot x_e^{\rm Saha}(z)}
{D_1}.
\eeq
At the transition redshift $z_{\rm t}=1570$, we use $x_e^{\rm corr}(z_{\rm t})$ as an initial condition for the integration with our full ODE.

\bibliography{emla}

\end{document}